# The dynamics and flexibility of protein disulphide-isomerase (PDI): predictions of experimentally-observed domain motions

J Emilio Jimenez-Roldan[1], Moitrayee Bhattacharyya[2], Stephen A Wells[1], Rudolf A Römer[1,4], Saraswathi Vishweshwara[2] and Robert B Freedman[3]

[1] Department of Physics, The University of Warwick, Coventry CV4 7AL, UK

[2] Molecular Biophysics Unit, Indian Institute of Science, Bangalore 560012, India

[3] School of Life Sciences, The University of Warwick, Coventry CV4 7AL, UK

[4] Centre for Scientific Computing, The University of Warwick, Coventry CV4 7AL, UK

**Contact**: Rudolf A Römer, Email: r.roemer@warwick.ac.uk, Phone +44 92476)574327, FAX (+44) 2476 2476 150897

**Running Title**: The dynamics and flexibility of PDI

**Summary:** We have studied the mobility of the folding catalyst, protein disulphide-isomerase (PDI), by molecular dynamics and by a rapid approach based on flexibility. We analysed our simulations using measures of backbone movement, relative positions and orientations of domains, and distances between functional sites. Despite their different assumptions, the two methods are surprisingly consistent. Both methods agree that motion of domains is dominated by hinge and rotation motion of the **a** and **a'** domains relative to the central **b-b'** domain core. We identify the **a'** domain as showing the greatest intra-domain mobility. The flexibility method, which requires $10^4$-fold less computer power, predicts additional large-scale features of inter-domain motion that have been observed experimentally. We conclude that the methods offer complementary insight into the motion of this large protein and provide detailed structural models that characterise its functionally-significant conformational changes.

**Graphical Abstract:**

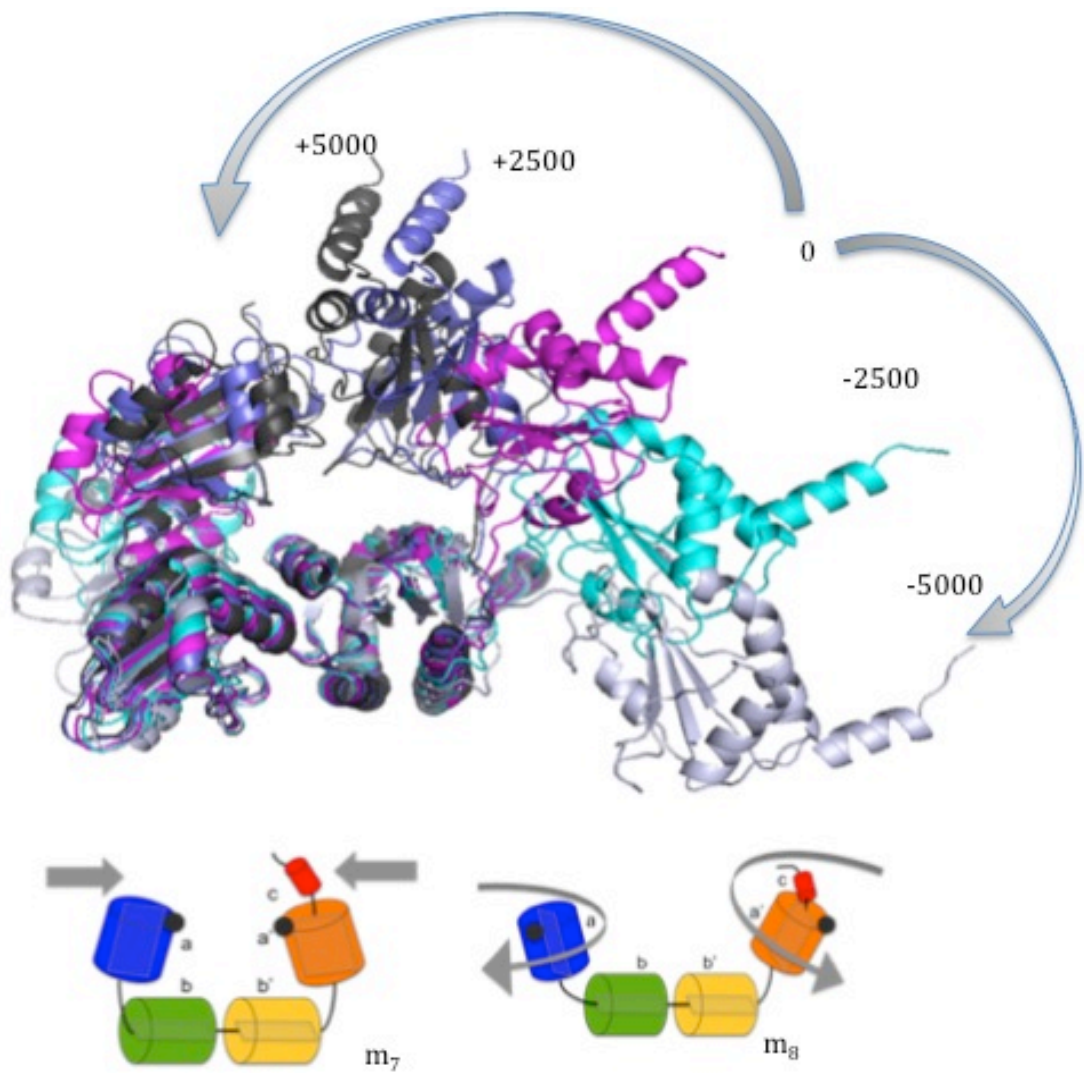

**Highlights:**

- Flexibility of yeast PDI has been explored by MD and a new rapid all-atom approach

- Both methods highlight PDI's capacity for notable inter-domain (twist/tilt) motion

- The rapid method predicts experimentally-observed extreme domain re-orientations

- Used in synergy, the methods can simulate fully the flexibility of large proteins

# Introduction

Proteins show internal mobility over a wide range of time- and length-scales, from the rapid local motions that define the conformational entropy of a given protein conformation (<ns), to the functionally-significant, concerted, slower motions (> ms) of loops or whole domains that interconvert different conformations (Henzler-Wildman and Kern 2007, Tzeng and Kalodimos 2012). However, there is a real limitation in our ability to describe fully the dynamics and flexibility of proteins and to characterize their motions – both in isolation and in response to interactions with partner proteins and other molecules. The gold standard approaches in this field are protein structure determination to atomic resolution by X-ray crystallography, and computationally-intensive simulations of motion by all-atom molecular dynamics (MD) calculations. For small proteins, MD simulations routinely extend over timescales up to 100 ns and can effectively identify local motions. But the most functionally-significant motions, involved in protein activity and protein:protein recognition, involve larger-scale structural units and motions – such as closure of a cleft between domains– which occur on timescales orders of magnitude longer than those accessible by typical MD simulations. There is a clear need for methods which can simulate protein motion that are less computationally-intensive but retain the full physical reality of the structure of the protein at the atomic level.

An attractive approach is to deconstruct and explore the protein structures in the form of networks of connections. Such networks at different levels (backbone, side chain and H-bonds based connectivity) have been investigated earlier (Thorpe, Lei et al. 2001, Greene and Higman 2003, Brinda and Vishveshwara 2005, Glembo, Farrell et al. 2012). We developed a consistent method for deconstructing a protein structure into a network of rigid and flexible units and then for exploring the natural modes of motion of this network, based solely on information contained within the high-resolution structure (Jimenez-Roldan, Freedman et al. 2012). The method can characterize plausible modes of motion in terms of the regions involved, the 'hinge' points, and can also explore the limits on the amplitude of motion associated with any given mode. We previously applied this method to 6 proteins from typical small 'model' proteins of <100 residues to large proteins with >1000 residues.

One of the large proteins investigated was protein disulphide-isomerase (PDI), an abundant catalyst of oxidative protein folding in the endoplasmic reticulum of secretory cells (cf. **Figure 1**). PDI is essential for the accurate and efficient co- and post-translational folding of secreted and cell-surface proteins including medically-important classes of protein such as antibodies, cytokines, digestive enzymes, blood-clotting factors etc.; indeed, manipulation of PDI levels is a successful strategy used in industry for increasing the yields of high-value recombinant disulphide-bonded proteins, see e.g. (Finnis, Payne et al. 2010). PDI comprises four domains, each of which has the conserved thioredoxin-fold conformation (trx-fold), arranged in sequence **a-b-b'-x-a'-c**, where **a** and **a'** are redox-active trx-fold domains, **b** and **b'** are structurally-similar domains lacking redox activity, **x** is an extended linker and **c** is a highly acidic C-terminal extension (Hatahet and Ruddock 2009) (refer to **Figure 1**a). The function of PDI within the cell involves successive interactions with a range of diverse partner proteins and protein substrates (Kozlov, Maattanen et al. 2010, Bulleid and Ellgaard 2011) which implies that PDI is a dynamic molecule, a conclusion supported by the long delay between its discovery in the 1960s and the first determination of a crystal structure of a full-length multi-domain PDI (Tian, Xiang et al. 2006). There is now a substantial body of experimental evidence supporting this view of PDI as a conformationally dynamic protein whose flexibility underlies its function. Studies on mammalian PDI, and also on PDI from the fungus *H.insolens*, indicate that there is extensive relative movement of the a' and b' domains, facilitated by the x-linker (see Discussion for further detail and references). By contrast, crystallization of yeast (*S.cerevisiae*) PDI (yPDI) at different temperatures produced crystal structures which differ markedly in the relative orientation of the **a** and **b** domains (Tian, Xiang et al. 2006, Tian, Kober et al. 2008).

In this paper we have three purposes: (i) to explore and describe in detail the motions of PDI, predicted by our novel approach, (ii) to compare this set of motions with those predicted by a gold-standard multi-ns all atom MD simulation and (iii) to compare the predictions of both methods to the experimental data on yPDI flexibility. As mentioned earlier, the rapid flexibility approach has accessed a large conformational space. In particular, starting from the crystal structure of yPDI determined at 4$^o$C, a conformation found in crystals grown at room temperature is accessed by the flexibility approach. This new conformation is not accessed by a relatively short (30 ns) MD simulation, highlighting the strength of the flexibility approach in rapidly exploring realistic regions of conformational space.

## Results

**Characterisation of yPDI motion predicted from rigid cluster and normal mode analyses**

The starting point for our approach to mobility simulation of full-length yPDI is the high-resolution structure (pdb: 2B5E), which is used to generate a representation of the molecule as a set of rigid clusters and flexible linkers. Rigidity analysis includes bonds as constraints and then proceeds by successive exclusion of the weakest bonds in the structure (Thorpe, Lei et al. 2001) (see Supplementary material and Figure S1). Our method for simulating protein motion (Jimenez-Roldan, Freedman et al. 2012) then applies normal mode analyses (Suhre and Sanejouand 2004, Dykeman and Sankey 2010) to the high-resolution structure to derive vectors that describe the low-frequency modes of motion of the protein. It then models the motion through a series of steps along the vectors (Wells, Menor et al. 2005, Jolley, Wells et al. 2008), treating clusters as rigid and other regions as flexible while maintaining constraints of local bonding geometry and steric exclusion. In this analysis, we have focussed on normal modes 7-11 ($m_7 - m_{11}$) which are the five lowest-frequency non-trivial normal modes ($m_1 - m_6$ describe trivial rotations and translations). We find that the motions are primarily motions of **a** and (especially) **a'** domains relative to a **bb'** base. **Figure 1b** summarises the trajectory along mode $m_7$ by showing superimposed structures representing the conformers computed at steps 0, + 1000, +2500, -1000 and -2500. The central domains **b** and **b'** form an essentially invariant base, while in comparison the N- and C-terminal domains **a** and **a'** show co-ordinated movement towards and away from each other, relative to their positions in the crystal structure; movement in the positive sense closes the structure, while that in the negative sense opens it. **Figure 1b** also represents this motion in cartoon terms and shows a similar representation of yPDI motion along the next lowest-frequency mode $m_8$ where the motions are essentially rotations of domains **a** and **a'** around intra-domain axes. For the next low-frequency modes $m_9$-$m_{11}$, the motion can be roughly characterised as comprising combinations of hinge motions and rotations of domains **a** and **a'**.

**MD study of yPDI highlights relative motion of domains**

In parallel with this analysis, we have undertaken an MD simulation study of the potential intramolecular motion of yPDI at 300K, based on the same high resolution x-ray structure 2B5E. We generated a full molecular trajectory over 30 ns (consuming >36,000 CPU hours) and have interrogated the trajectory in a number of ways in order to compare it with the outputs from the computationally less demanding flexibility approach. Both the flexibility and MD simulations generate trajectories comprising a series of all-atom protein structures and hence both can be subjected to the same analyses.

Initially we asked whether we could confirm that the major features of motion were motions of whole domains. To test if the domains remained essentially intact as structural units, we obtained measures of β-sheet geometry for each domain and asked how constant these measures remained through the simulations of motion. For each domain, we selected a number of $C_\alpha$ atoms to define the sheet and extracted angles between sets of three atoms as measures of sheet geometry. We then derived the mean and standard deviation of these angles through the trajectories of motion obtained both by MD simulation and flexibility analysis. **Figure 1c** plots these data for each domain and shows (i) that each angle measure shows a very narrow variation through the motion simulations (standard deviations are +/- 2-3° for the flexibility simulations and +/-5-6° for the MD simulations), and (ii) that there is very close agreement between both types of simulation in the mean angles observed. These data confirm that the core β-sheets of each domain retain essentially constant geometry throughout both sets of simulation (see also Figure S2).

**Analysis of inter-domain motion predicted by flexibility and MD simulations**

In order to provide a quantitative account of the large-scale *inter-domain* motion of this multi-domain protein, we have represented each domain as a plane based on its core *β*-sheet and calculated the relative orientations of these planes (tilt (θ) and dihedral twist (δ)) through simulations of motion. **Figure 2** presents a twist/tilt plot for each adjacent domain pair and allows a direct comparison of the predictions from flexibility analysis and MD simulations. The upper panel represents the **a-b** domain pair and shows that the MD simulation explores a considerable range around the 2B5E starting structure, but that all orientations fall close to a single line in twist/tilt space. Flexibility analyses in modes $m_7$ and $m_8$ predict the same pattern of motion, but modes $m_9$ and $m_{10}$ begin to explore other regions of twist/tilt space. In particular, motion along the negative direction of $m_{10}$, converts the relative orientation of the **a** and **b** domains into precisely that found in the alternative 'high-temperature' crystal structure of yeast PDI (pdb: 3B0A); this structure lies far away from the region explored in the MD simulation.

The central panel shows comparable data for the **b-b'** domain pair and indicates that MD and flexibility both explore the same limited region of twist/tilt space but with the flexibility simulation extending slightly further away from the orientations defined by the crystal structures. The data in the lower panel refer to the **b'-a'** domain pair and show different characteristics. The MD simulation covers an extended area of twist/tilt space while the normal mode trajectories form various paths through this area and also extend well beyond it; both simulations predict very extensive relative motion of this domain pair. (See also **Figure S3**)

To indicate the scale of the relative motion of domains in our simulations, we have used as a measure the distance between active site residues located in domains **a** (Cys61 in the full-length sequence) and **a'** (Cys406), respectively. This distance is 27Å in the crystal structure, but **Figure 3a** demonstrates that movement along $m_7$ allows it to vary freely from 15 – 80Å due to the co-ordinated hinge-like motion of domains **a** and **a'** relative to **b-b'**. **Figure 3a** also shows the less striking evolution of the inter-active-site distance through the other normal mode trajectories, reflecting the more complex motions in these modes. **Figure 3b** shows that during the 30 ns MD simulation, the inter-active-site distance increases from its initial value and then varies widely through the trajectory, ranging up to 68Å. (See also **Figure S4** and **Figure S5**).

**Analysis of intra-domain motion predicted by flexibility and MD simulations**

To focus more on local, *intra-domain* motion, we have determined the pseudodihedral angle ξ (Warshel and Levitt 1976, DeWitte and Shakhnovich 1994) at each residue for each structure computed in modes $m_7$-$m_{10}$ and then derived the RMS variation of cos ξ, as a measure of the net flexibility at each residue. As shown in **Figure 4a**, this analysis highlights as regions of greatest flexibility the N- and C-termini and the **x**-linker, with local maxima also shown at the **a-b** and **b-b'** domain boundaries. This confirms that the major flexibility of the protein arises from the relative motion of domains but it also indicates that there is considerable intra-domain motion. **Figure 4a** also displays the variation of pseudodihedral angle ξ, as derived from the MD analysis. MD detects much more local and higher frequency motion and hence it is not surprising that the MD-derived plot shows much greater scatter. Taking a running average over 5 residues smooths some of this and again shows that the most marked motion is at the termini and at the **x**-linker; motion at the other domain boundaries is not more marked than at several intra-domain sites. Interestingly, both methods indicate much greater intra-domain motion within the **a'** domain than within the other domains.

To obtain an alternative measure of the extent of intra-domain motion through our simulations, we monitored the evolution of the MD trajectory over 30 ns in terms of the RMSD between each of the domains and the structure of that domain in the initial full-length protein (**Figure 4b**). It is apparent that each domain initially deviates to some extent from its original structure in the crystal, but that for the period from 5-25 ns, these differences are approximately stable, with domain **a'** showing the greatest RMSD and domains **a** and **b** showing the smallest. What also emerges from **Figure 4b** is that the MD analysis over this period does not explore the full potential for motion. In the period from 25 nsec onwards, domain **b** (in contrast to the other domains) begins a further phase of motion, which considerably increases its RMSD from the starting structure. For comparison, we analysed the change in RMSD for each domain through the flexibility trajectories of modes $m_7$-$m_{11}$. (see also Figure S6).

**MD analysis of extreme 'closed' structure generated by flexibility analysis**

It is noteworthy that the MD trajectory (**Figure 3b**) shows the distances between active-site Cys residues (values range from 22 to 68 Å) as typically much greater than the initial value (~26 Å) and does not explore more 'closed' structures. By contrast, our analysis of motion along normal mode $m_7$ suggested that the molecule was able both to 'open' and 'close' relative to the starting structure, generating values for the inter-active site distance in the range 15-80 Å (**Figure 3a**). To test whether such closed structures were artefacts of our flexibility approach, a short MD simulation (10 ns) was performed on a 'closed' structure to assess its physical plausibility. We took the atomic co-ordinates of a 'closed' structure representing the extreme of positive motion along $m_7$ and used it as the starting point for the MD simulation. **Figure 5a** shows the MD trajectory over 10 ns starting from this extreme 'closed' structure, expressed in terms of the inter-active-site distance. The molecule moves gradually to explore structures in which the distance between the active site residues 61 and 406 lies in the range from 8 – 17 Å. The intra-domain distance between cysteines at residues 61 and 90 remains constant during the simulation as in **Figure 3b**. The inter-cysteine distance for the 90-406 pair is correctly coordinated with the 61-406 pair. Two conclusions are very clear from this simulation, (i) the closed structure is physically plausible as it is not immediately abandoned in the first few ns of the MD simulation, and an ensemble of 'closed' conformations are explored by MD simulations, (ii) the original 30 ns MD simulation did not fully explore the conformational space available to yPDI, since it never generated any structure comparable to those found in this 10 ns simulation (as judged by the value of the inter-site distance). These conclusions are confirmed by the data in **Figure 5b,** which represent the trajectories from the 10 ns MD simulation in terms of twist/tilt plots. Hence the flexibility analysis clearly finds energetically plausible inter-domain orientations, which do not simply regress from the 'closed' structure towards the orientations found in the crystal structure. It is known in the literature (Caves, Evanseck et al. 1998, Loccisano, Acevedo et al. 2004) that short multiple MD simulations, with different starting conformations, can explore the conformational space better than a single long equilibrium simulation. Our results from two MD simulations reiterate this point. More importantly, we have shown that the flexibility approach can complement MD simulations by quickly providing such a set of realistic, different starting points.

## Discussion

The main feature of PDI molecular motion predicted by both flexibility and MD approaches is that the redox-active **a** and **a'** domains show considerable freedom of motion with respect to a relatively fixed base provided by the **b** and **b'** domains. This motion includes both rotations and hinge-like opening and closing about the **a-b** and **b'-a'** hinges. These findings are valuable in reconciling discrepancies in data from literature on the mobilities of PDIs from the yeast *S.cerevisiae* and other sources. A substantial body of work on human PDI and its fragments in solution, using NMR and intrinsic fluorescence, has identified motion of the **x**-linker region relative to the adjacent **b'** domain as a key feature; such motion would modulate ligand access to the major non-covalent ligand binding site on the **b'** domain and alter the orientation of this domain relative to the catalytic **a'** domain (Nguyen, Wallis et al. 2008, Byrne, Sidhu et al. 2009, Wang, Li et al. 2012). This is consistent with selective proteolysis studies on bovine and human PDI (Freedman, Gane et al. 1998, Wang, Chen et al. 2010), which showed preferential cleavage around the **x** linker indicative of substantial flexibility of the protein in the region between the **b'** and **a'** domains. A major change in relative orientation of these two domains was also inferred from NMR, x-ray and SAXS studies on the reduced and oxidised states of PDI from the fungus *H.insolens* (Nakasako, Maeno et al. 2010, Serve, Kamiya et al. 2010). A similar redox-driven re-orientation has recently been demonstrated in x-ray crystallography structures of truncated human PDI (Wang, Li et al. 2012). By contrast, work on yPDI has provided highly valuable insights, based on x-ray crystallographic structures, but little comparable work on mobility in solution. The x-ray work, based on yPDI crystals grown at different temperatures, has provided two structures which differ very little in the relative orientation of the **b'** and **a'** domains, but show clear differences in the N-terminal half of the protein, in the relative orientation of the **a** and **b** domains (Tian, Xiang et al. 2006, Tian, Kober et al. 2008). The simulations reported here are based on the higher resolution yeast structure, derived from crystals grown at 4º C. They show that there is extensive motion of both **a** and **a'** domains, with the motion of the **a'** domain being more marked, but that the mode 10 trajectory involves a rotation of the **a** domain relative to the **b** domain which results in the orientation found in the higher temperature structure. It is likely that yPDI in solution shows extensive mobility both in the **a-b** and in the **b'-x-a'** regions, but that the non-statistical sampling of conformations provided by crystallization has highlighted only the **a-b** motion.

Our rapid flexibility analysis requires only a few CPU hours to generate many trajectories for this large (c. 500 residues) protein in full atomistic detail, providing the basis for a close analysis of molecular motion. Given the fundamental conceptual and operational differences between MD and the rapid flexibility approach – and the difference of several orders of magnitude ($\times 10^4$) in computer time that MD requires – their predictions are strikingly congruent in many respects. The methods provide very similar predictions of the range of inter-domain orientations explored (**Figure 2**), the variation of inter-active-site distance that is accessible (**Figure 3**), and the extent to which the individual domains show intra-domain flexibility (**Figure 4**). The MD approach provides a far more detailed picture of local motion (**Figure 4a**) but does not explore some very large amplitude domain motions, which are detected by the flexibility approach and are supported by experimental data. One example is the very large change in the **a-b** inter-domain twist angle (**Figure 2**a) predicted by motion along mode 10, which generates precisely the relative orientation of these domains found in an alternative crystal structure (Tian, Kober et al. 2008). A further example is given by flexible motion along mode 7, which can generate 'closed' structures with a much shorter inter-active-site distance (ca. 15Å) than that found in either crystal structure. Experimental data indicate that these sites can be cross-linked by bifunctional chemical reagents with maximum span 16Å (Hawkins, de Nardi et al. 1991) highlighting the extent to which the 'horseshoe' structure of PDI can close in solution. Furthermore, although such structures were not generated in the initial 30 ns MD simulation, such closed structures are rather stable (**Figure 5**) over a 10 ns MD simulation, with limited amplitude fluctuations in domain orientation and inter-active-site distance.

In conclusion, the wide-ranging and disparate experimental data that indicate the extensive and functionally-significant flexible motion of PDI have now been supplemented with structurally-detailed simulations which confirm the character and extent of this flexibility. Both MD and the rapid flexibility approach predict the extensive relative domain motions, especially the motions of the **a'** and **a** domains relative to the **b-b'** core. The two approaches are complementary in that MD provides data on high-frequency local motion while the flexibility approach allows a more rapid exploration of the full range of inter-domain motion. Our comparison of the flexibility results with experimental data and with MD results gives us confidence to proceed to a more wide-ranging application of this approach to the many other PDI family structures that have been determined since 2006 when the yeast PDI (2B5E) structure was published (Kozlov, Maattanen et al. 2010).

## Experimental Procedures

**Flexibility simulation:** We are using a recently developed rapid method (Jimenez-Roldan, Freedman et al. 2012), combining protein rigidity analysis (Jacobs, Rader et al. 2001), geometric modelling of flexible motion (Wells, Menor et al. 2005, Jolley, Wells et al. 2006) and elastic network modelling (Suhre and Sanejouand 2004). This method moves the structure while maintaining bonding and steric constraints. Coarse-grained elastic network modelling identifies low-frequency modes, which are possible directions for flexible motion. Rigidity analysis rapidly identifies rigid clusters and flexible regions which can act as "hinges". These flexibility simulations allow us to explore large-amplitude motion along multiple normal modes in an all-atom protein structure at minimal computational expense (Jimenez-Roldan, Freedman et al. 2012) and provide valuable information on the flexibility of the protein and its functional sites. We have performed the analysis through multiple (2000-5000) steps in each direction from the starting structure, representing the full structure after every 100 steps; hence the analysis is always based on 10 trajectories (one in each direction for each of 5 normal modes) where each trajectory comprises 20-50 all-atom structures in the form of PDB files. We previously showed that each trajectory leads to a gradual shift of the protein from the starting structure, as measured by RMSD between original and derived structures, and that this shift may reach an asymptote, where no further motion is possible along the initial vector, as a result of steric constraints (Jimenez-Roldan, Freedman et al. 2012). See Supplemental Information for more details.

**Molecular dynamics:** All-atom MD simulations on yPDI were performed at 300K using the AMBER9 suit of programs (Case, Darden et al. 2006) with *ff03* force field (Duan, Wu et al. 2003) and parm99 parameters. Explicit solvent MD simulations are carried out on the 2B5E crystal structure of yPDI (for 30 ns) and a 'closed' structure of PDI (for 10 ns) generated from the geometric simulations as described above. The MD simulations are performed in aqueous medium using the TIP3P water model. The solvation box is 12Å from the farthest atom along any axis. $Na^+$ ions are added to neutralize the net charges on yPDI using the tleap module in AMBER9. The MD simulations are performed under NPT conditions using the Berendsen thermostat and periodic boundary conditions. Particle Mesh Ewald (PME) summation is used for long-range electrostatics and the van der Waals cut-off used is 10 Å. The pressure and temperature relaxations are set to 0.5 $ps^{-1}$. SHAKE constraints are applied to all bonds involving H atoms. A time step of 2 fs is employed with the integration algorithm and the structures are stored every 1 ps. All the MD simulations are implemented and analyzed using a 264 core Intel Xeon HPC cluster.

**Pseudodihedral RMS:** To generate the pseudodihedral $\xi_i$ for residue $i$ at $C_a$-atom position $\mathbf{r}_i$, we consider the inter-residue pseudo-bonds $\mathbf{q}_{i-1}$, $\mathbf{q}_i$, $\mathbf{q}_{i+1}$ defined such that $\mathbf{q}_i = \mathbf{r}_{i+1} - \mathbf{r}_i$ for all residues $i$ along the protein main chain (except for the first residue in the sequence and for the last two residues). The dihedral angle $\xi_i$ formed by $\mathbf{q}_{i-1}$ and $\mathbf{q}_{i+1}$ about the axis of $\mathbf{q}_i$ then characterizes the orientation of the residues and their variation and flexibility. When the protein main chain is in a fully extended (β-strand) state, $\xi_i$ is near zero ($\cos(\xi_i) \approx 1$). When the main chain is in a coiled or turned conformation (α-helix or β-hairpin), $\xi_i$ has magnitude above 90 degrees, and $\cos(\xi_i)$ is negative, typically around -0.7 for α-helices. We extract $\cos(\xi_i)$ for each residue $i$ and each conformer generated during a motion. We then use the root-mean-square-deviation of all such $\cos(\xi_i)$ values as our measure of flexibility at residue $i$.

**Tilt and dihedral twist:** Each of the **a**, **b**, **b'**, **a'** domains contains a central β-sheet, characteristic of the thioredoxin fold. The orientation of the domain can be represented by a vector normal to a central part of the β-sheet. This vector is generated by selecting the $C_\alpha$ atoms of four 'central' residues, namely alternating central residues in each of the adjacent anti-parallel strands $β_2$ and $β_4$ (by PDI numbering; these are strands 1 and 3 of the thioredoxin fold). These 4 atoms define a quadrilateral and from its two diagonals we construct a normal vector **n**. When we now consider two adjacent domains in the yPDI structure – labelled as domains 1 and 2, with central positions $\mathbf{r}_1$, $\mathbf{r}_2$ and plane normals $\mathbf{n}_1$, $\mathbf{n}_2$ – we can define an inter-domain tilt angle $\theta$ in the range 0 to 180 degrees as $\cos(\theta) = \mathbf{n}_1 \cdot \mathbf{n}_2$. An inter-domain dihedral twist $\delta$, with range from -180 to +180 degrees, is obtained by constructing an inter-plane vector $\mathbf{r}_{12}=\mathbf{r}_2-\mathbf{r}_1$ and considering the dihedral $\delta$ between the plane containing $\mathbf{n}_1$, $\mathbf{r}_{12}$ and the plane containing $\mathbf{n}_2$, $\mathbf{r}_{12}$ (see Supplemental Information).

**Acknowledgements**

We thank Jack Heal and Mike Thorpe for inspiring discussions. SA Wells gratefully acknowledges the Leverhulme Trust for an Early Career Fellowship. We are grateful to the Royal Society and the Indian DST for support of a UKIERI Scientific Seminar and also the Warwick Institute of Advanced Study for funding the "Protein Biology and BioPhysics" network as well as the EPSRC MOAC DTC for providing facilities. We thank the DBT, Govt. of India, for computational support for the MD simulations. S Vishweshwara acknowledges a CSIR emeritus scientist fellowship. EJR, RBF, RAR and SAW planned the study. EJR and SAW ran the flexibility analysis; MB and SV computed the MD trajectories. All authors performed data analysis and discussed the results. RBF and RAR wrote the manuscript while EJR, MB, and SAW supplied the figure data.

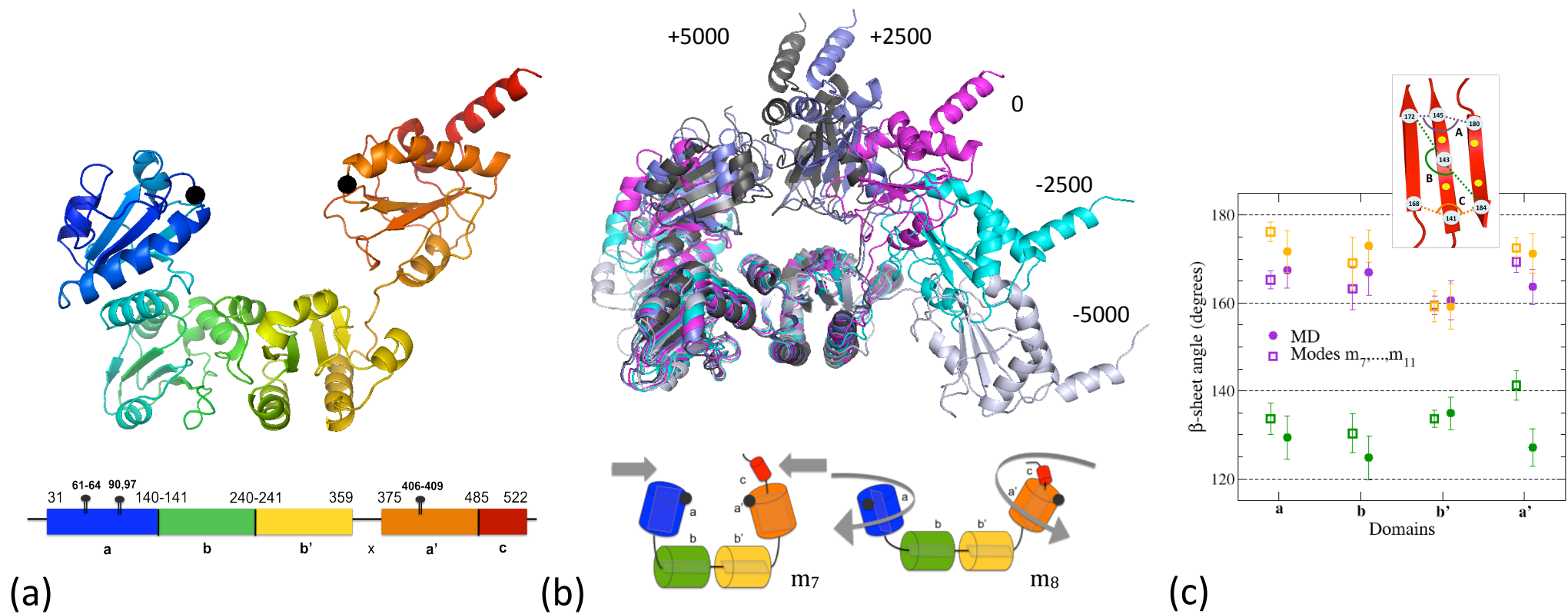

**Figure 1** The structure of yPDI (2B5E) and its inter-domain motion (a) Schematic ribbon diagram (centre) and cartoon (top) of the tertiary structure of yPDI coloured from N-terminus to C-terminus with the **a** domain predominantly blue, the **b** domain green, the **b'** domain yellow, and the **a'** domain orange. The C-terminal is shown in red and the α-carbons of the active site cysteine residues 61 and 406 are shown as black spheres.. The flexible **x** region is given between domains **b'** and **a'** in light orange. Domains **a**, **b'** and **a'** are situated in the same spatial plane but domain **b** (green) is displaced away from the reader. The thick line at the bottom shows the domain organisation of yPDI based on the crystal structure with the four domains (**a-b-b'-a'**), the flexible loop **x** connecting domains **b'** and **a'** and the C-terminal tail **c**. Cysteine residues are shown as stalks and numbered. (b) Snapshots of conformational change for yPDI moving along mode $m_7$. The figure shows the overlap of conformers 0, +2500, +5000, -2500, -5000, aligned on domains **b** and **b'**. At the bottom, we give a cartoon representation of conformational motion in normal modes $m_7$ and $m_8$. The planes in each domain and the two black circles schematically indicate the β-sheets and the two active sites. (c) Averages for the relative β-sheet angles in all four domains as computed from the 30 ns MD (closed symbols) and also the flexibility analysis (open symbols, averaged over modes $m_7, \ldots, m_{11}$). The three different colors indicate the three measured angles for the residue triples as given in the schematic, e.g. for the **b** domain with residue numbers as shown.

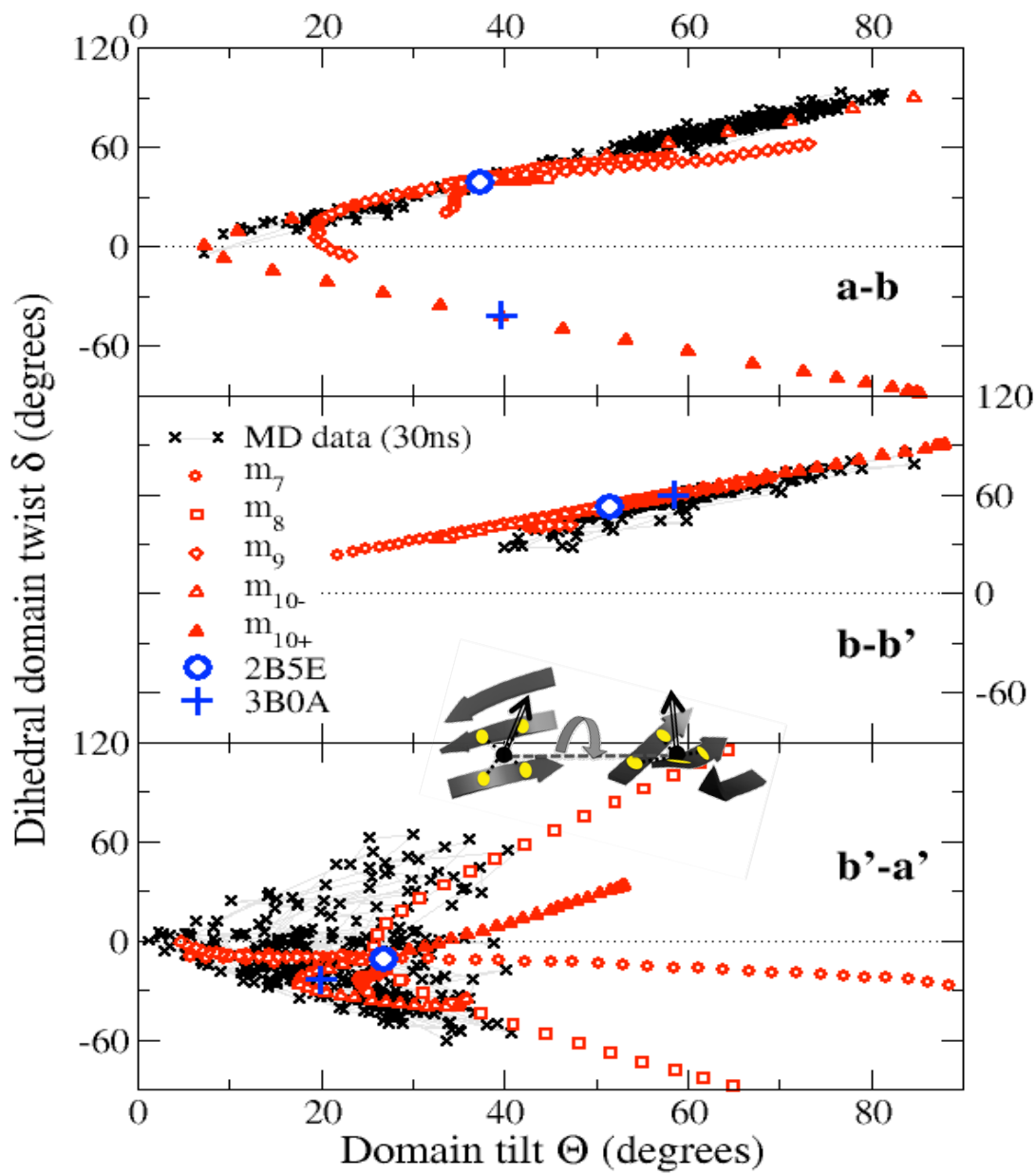

**Figure 2** Characterization of interdomain mobility of yPDI. The panels show the tilt and dihedral twist motion (see **Error! Reference source not found.**) between neighbouring domain pairs **a-b**, **b-b'** and **b'-a'**. The black crosses indicate results from the MD run, whereas the red symbols denote the modes $m_7$, $m_8$, $m_9$ and $m_{10}$. For $m_{10}$, we additionally distinguish between the directions of motion $m_{10-}$ and $m_{10+}$. The blue symbols denote tilt and twist for the two indicated crystal structures. The thin grey lines connect the MD results as they are computed along the MD trajectory. The horizontal dotted lines indicate a twist $\delta=0$. The inset graphics in the central panel indicates schematically the geometric interpretation of tilt and twist angles based on a quadrilateral anchored at 4 residues (yellow spots) on each $\beta$-sheet (see also Supplemental Information).

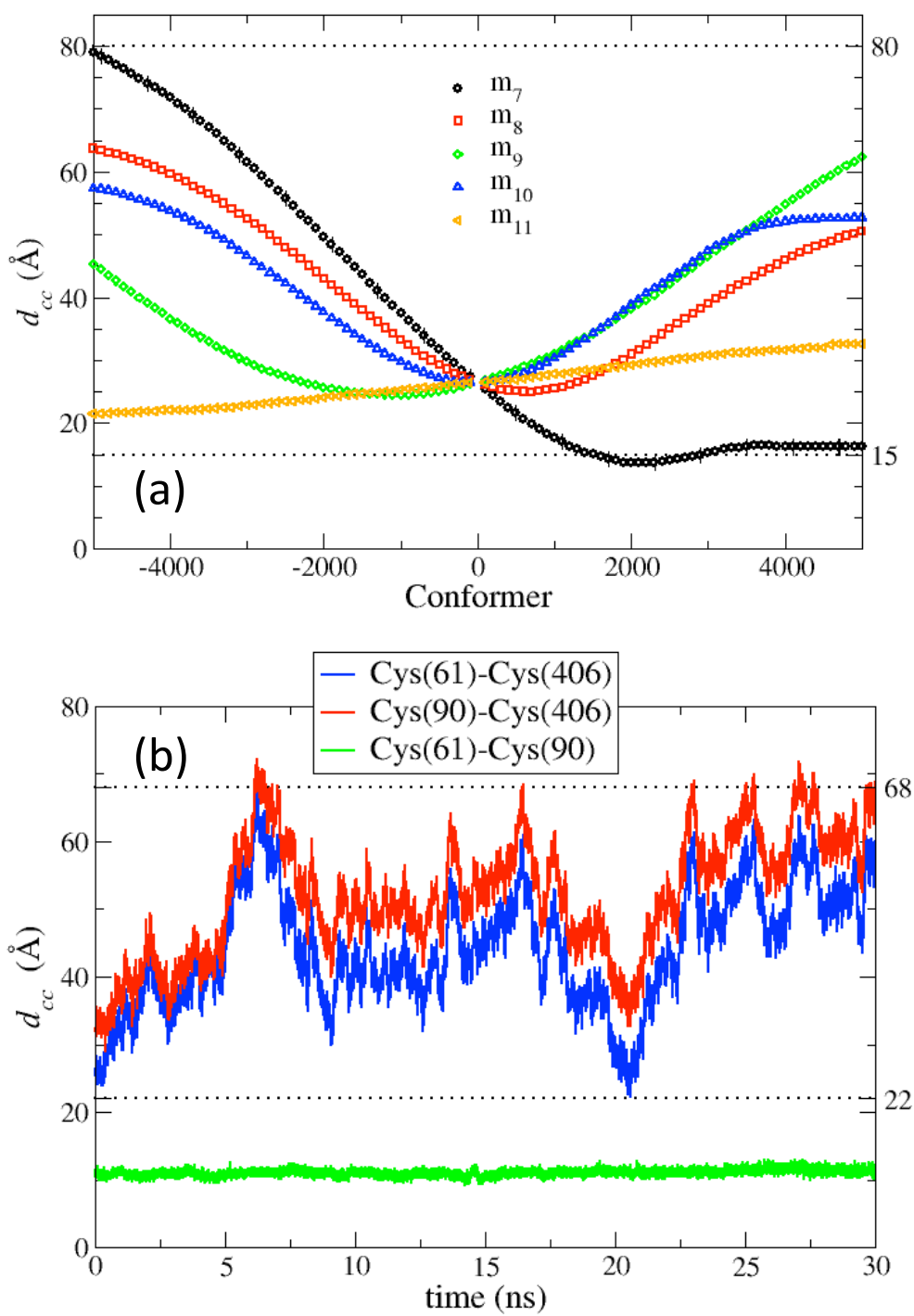

**Figure 3** Interdomain motion generates large change in distance between active sites. (a) Distance $d_{cc}$ between the $C_\alpha$ atoms of cysteine residues Cys61 and Cys406 as a function of conformer generated as the structure moves along the normal modes. Error bars indicate error-of-mean for 5 different starting conditions (shown for $m_7$ and every $5^{th}$ symbol only). (b) Evolution of $d_{cc}$ between cysteine pairs for the 30ns MD simulation. The horizontal dotted lines indicate the observed range of distances for the Cys61-Cys406 pair.

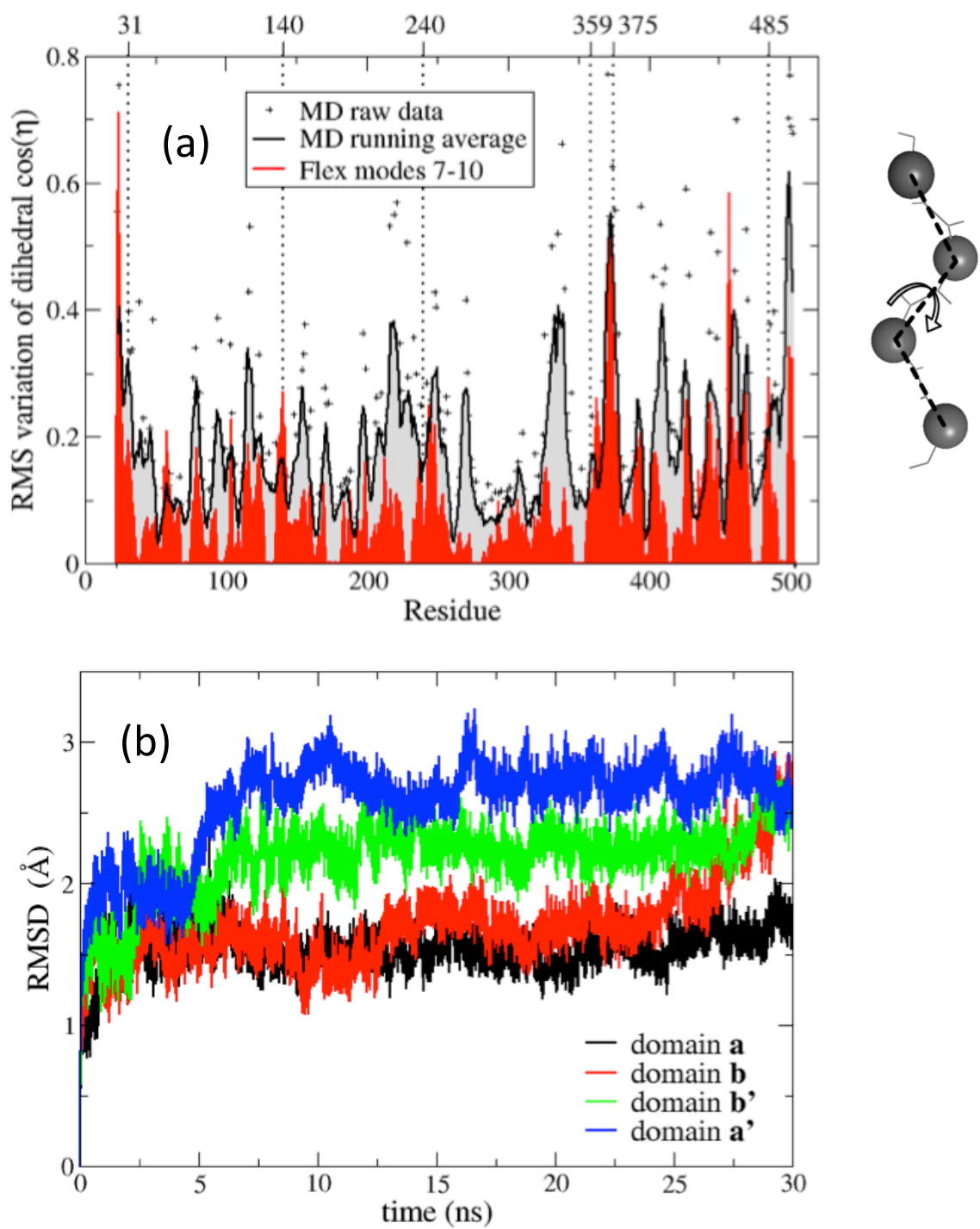

**Figure 4** Domains show clear differences in intra-domain mobility (a) Pseudodihedral RMS variation plot vs. sequence from long MD run (raw data and running average over 5 residues) and for the flexibility analysis averaged over $m_7, \ldots, m_{10}$ at $E_{cut}$ = -2 kcal/mol. The domain boundaries are marked with dotted vertical lines. (b) RMSD as a function of simulation time for yPDI domains. The values are obtained by overlapping each domain from the initial crystal structure with itself from the conformers generated during a 30ns MD simulation.

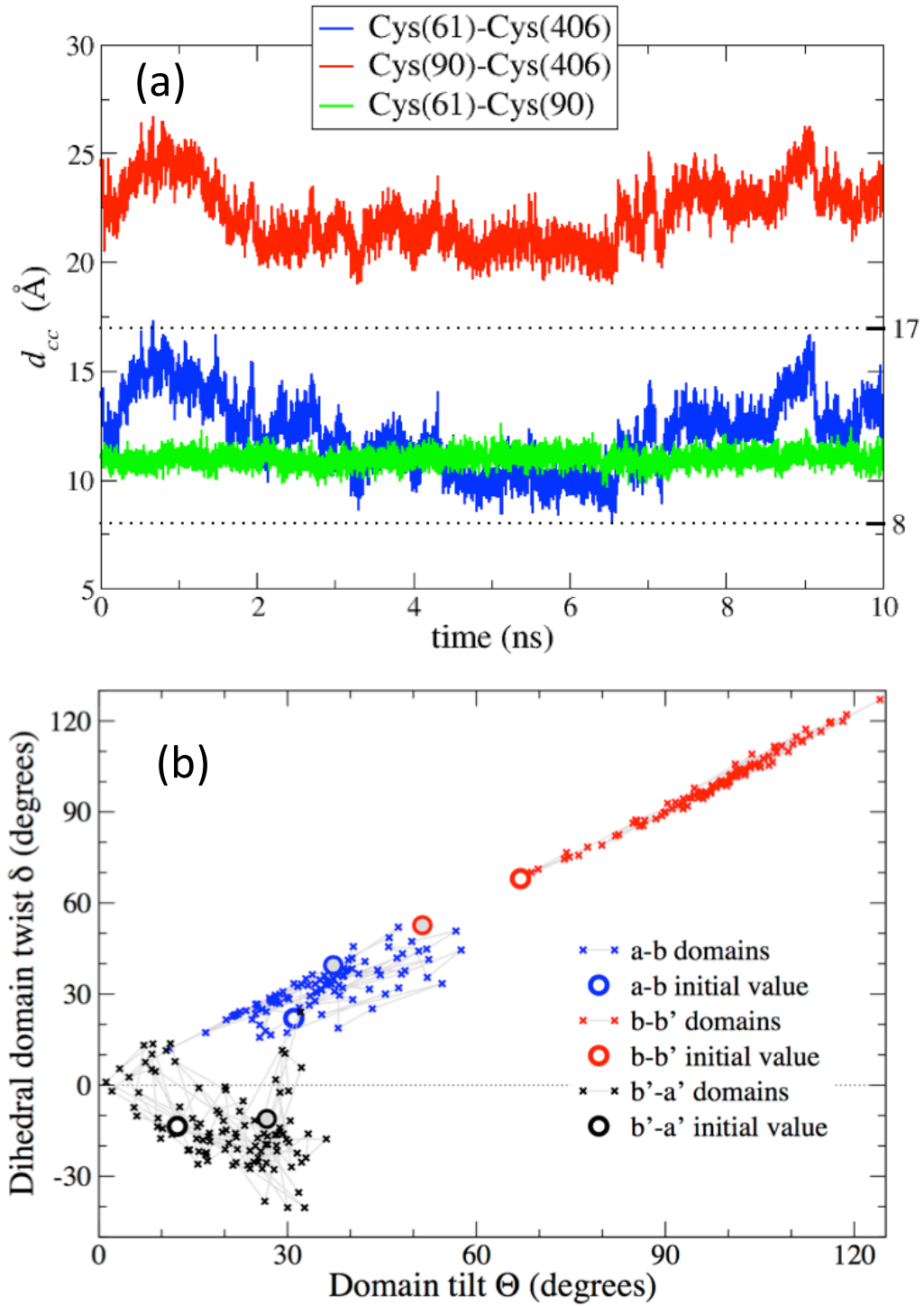

**Figure 5** MD analysis of extreme structure generated by flexibility shows that it is stable (a) Evolution of inter-cysteine distances between cysteine pairs for the 10ns simulation. (b) Tilt/twist analysis of the short MD run. Here the initial tilt/twist values (circles) are from the closed structure found via the flexibility analysis. The grey-shaded initial tilt/twist values correspond to the crystal structure of 2B5E.

# Supplemental Information for "The dynamics and flexibility of protein disulphide-isomerase (PDI): simulations predict experimentally-observed domain motions"

JE Jimenez-Roldan[1], M Bhattacharyya[2], SA Wells[1], RA Römer[1,4], S Vishweshwara[2] and RB Freedman[3]

### Rigid cluster analysis of yeast PDI

The starting point for our approach to mobility simulation of full-length yeast PDI is the high-resolution structure (pdb: 2B5E), which is used to generate a representation of the molecule as a set of rigid clusters and flexible linkers. In **Figure 1** we show the structure schematically both as a ribbon diagram and as a linear representation (), highlighting its organization as 4 distinct domains with an extended linker between domains **b'** and **a'**. The rigid cluster analysis includes bonds as constraints but is dependent on setting a 'cut-off' energy parameter ($E_{cut}$) to include strong H-bonds and exclude weaker ones (as inferred from the high-resolution structure). **Figure S1** illustrates the effect of varying the value of $E_{cut}$ on the analysis. Inclusion of all H-bonds leads to a representation of PDI as a single rigid cluster. Exclusion of the weakest bonds gives a model that shows the molecule as four distinct rigid clusters corresponding closely to the structurally-defined domains. Further stepwise exclusion of weaker bonds then leads either to fragmentation of each cluster into a series of smaller clusters (see e.g. domain **a'**) or to maintenance of the domain as a single cluster, but with specific regions becoming flexible and hence excluded from the cluster (see e.g. domain **a**). We have performed the rigidity and mobility analyses at several values of $E_{cut}$ and found that, in practice, the nature of the results is not very sensitive to $E_{cut}$ (as shown in **Figure S1**. For the more detailed analysis in the paper, we have focussed on simulation at $E_{cut}$ = -2 kcal/mol; previous work on other proteins has shown that this value makes flexibility predictions that are borne out by unexpected experimental findings (Li, Wells et al. 2012, Amin, Wallis et al. 2013). Also, additional modes could be explored, and we have analysed trajectories for modes $m_{12}$-$m_{16}$ but we find that the lowest 5 modes capture the majority of the accessible motion and provide a good overall picture with considerable structural detail. In many cases the mode trajectory is eventually limited by steric constraints, providing an amplitude limit on the motion.

## More details on the mobility analysis

The measures of β–sheet angles as shown in **Figure 1** represent averages over the 30ns MD. **Figure S2** shows the full evolution of these β–sheet angles through the MD simulation, demonstrating the limited range of values found for each angle and the lack of trend over time. This result indicates the remarkable stability of the β–sheet angles and hence of the integrity of domains during motion in yPDI.

An alternative analysis of the relative motion of domains can be made by selecting a single central point to represent each domain and then calculating through each trajectory the four-domain dihedral angle (**a-b** vs. **b'-a'**) and the two inter-domain angles **a-b-b'** and **b-b'-a'.** These measures capture the relative overall positions of domains, but not their relative orientations. As presented in **Figure S3**, this analysis shows that MD simulations and flexibility simulations make similar predictions for the range covered by the **b-b'-a'** inter-domain angle and come close to predicting the relative domain positions found in the 'alternative' crystal structure of yeast PDI. But the simulations differ in respect of the range of **a-b-b'** angle and the range of four-domain dihedral angle found; in both cases, the flexibility simulations explore a wider range.

## Comparing MD and the flexibility approach

Empirical-potential molecular dynamics and rapid simulations of flexible motion have quite different conceptual bases. MD makes use of a detailed force-field which defines a high-dimensional energy landscape. Physically acceptable conformations of the protein correspond to low-energy regions of this landscape, while unacceptable conformations have high energy. The MD simulation moves across the energy landscape by numerical integration of classical equations of motion. In contrast, the flexibility simulations place constraints on local bonding geometry and steric exclusion while neglecting longer-range interactions. This has the effect of simplifying and flattening the low-energy regions of the MD energy landscape, removing a large number of small energetic or kinetic barriers, while still forbidding access to the high-energy regions with unacceptable steric overlaps or distortions of bonding geometry. The flexibility simulations can pass rapidly between different regions in conformational space, which are in principle accessible to MD simulations but in practice would only be explored on very long timescales. There are several methods including replica exchange MD (REMD), umbrella sampling, etc. (Liwo, Czaplewski et al. 2008) to sample a wider conformational space.

# More on the use of the structural measures

**Pseudodihedral RMS:**

The conformation of the protein main chain can be almost completely specified by giving the two Ramachandran angles ($\varphi,\psi$) for each residue. These define the geometry of the C-N-$C_\alpha$-C and N-$C_\alpha$-C-N variable dihedrals around the $C_\alpha$ atom. The variation of the Ramachandran angles in the course of a simulation is then a measure of the flexibility of the protein main chain. However, much of the information present in the Ramachandran angles can be captured using one number per residue rather than two, by considering a pseudodihedral measure defined by the $C_\alpha$ atoms of four successive residues (Levitt 1976, DeWitte and Shakhnovich 1994). The variation of this backbone pseudodihedral is, in turn, a convenient measure of protein flexibility.

Obviously we are concerned less with the absolute value of $\xi_i$, but more with the variation in $\xi_i$ during motion. Our approach to describing the flexibility of the protein is as follows. For each structure generated during simulations of motion, we extract $\cos(\xi_i)$ for each residue $i$. We then find the mean and variance of $\cos(\xi_i)$ over the course of the simulation for each residue. The root-mean-square-deviation of $\cos(\xi_i)$ is our measure of flexibility.

**Tilt and dihedral twist:**

Let us identify the atoms at the corners of a quadrilateral as a, b, c, d in cyclic order (cp. the schematic in **Figure 2**). They have position vectors $\mathbf{r}_a$, $\mathbf{r}_b$, $\mathbf{r}_c$, $\mathbf{r}_d$. We obtain a central position $\mathbf{r}$ for the plane as $\mathbf{r}=(\mathbf{r}_a+\mathbf{r}_b+\mathbf{r}_c+\mathbf{r}_d)/4$. Vectors representing the diagonals of the quadrilateral are $\mathbf{r}_{ac}=\mathbf{r}_c-\mathbf{r}_a$ and $\mathbf{r}_{bd}=\mathbf{r}_d-\mathbf{r}_b$ and $\mathbf{n}= \mathbf{r}_{ac}\times\mathbf{r}_{bd}$ is the unit normal vector $\mathbf{n}$ parallel to the cross product of the two.

We now consider two adjacent domains in the yPDI structure, which we label as domains 1 and 2, with central positions $\mathbf{r}_1$, $\mathbf{r}_2$ and plane normals $\mathbf{n}_1$, $\mathbf{n}_2$. An interdomain tilt angle $\theta$ in the range 0 to 180 degrees is obtained from $\cos(\theta)=\mathbf{n}_1\cdot\mathbf{n}_2$. An interdomain dihedral twist $\delta$ in the range -180 to +180 degrees is obtained by constructing an interplane vector $\mathbf{r}_{12}=\mathbf{r}_2-\mathbf{r}_1$ and considering the dihedral $\delta$ between the plane containing $\mathbf{n}_1$, $\mathbf{r}_{12}$ and the plane containing $\mathbf{n}_2$, $\mathbf{r}_{12}$. Two natural motions for adjacent domains are (i) a "towards-and-away" tilting motion, in which $\theta$ varies substantially with little variation in $\delta$, and (ii) an "axial twist" motion of rotation about the interplane vector, generating a co-variation of $\theta$ and $\delta$. Motions of this kind are visible in our tilt-twist plots as horizontal and diagonal trajectories.

Tilt and twist values were extracted for the input structures, for the structures generated in the simulations of flexible motion, and for conformers of the MD trajectory, to describe three relative domain orientations: **a-b**, **b-b'**, and **b'-a'**. The residues selected to represent plane orientations in yPDI are as follows: In domain **a**, residues LEU 108, ILE 110, LEU 53, and GLU 55; in domain **b**, residues LEU 202, ILE 204, ILE 163 and GLN 165; in domain **b'**, residues PHE 314, ILE 316, GLY 260 and LEU 262; last, in domain **a'**, residues ILE 453, LEU 455, LEU 398 and LEU 400. Note that we use next-nearest-neighbour rather than nearest-neighbour residues in each strand, in order to prevent the pleating of the $\beta$-sheet affecting our results.

## Additional analysis of the inter- and intra-domain motion

**Figure 3** shows how flexibility (**Figure 3**a) and MD (**Figure 3**b) explore the inter-domain motion. Extending this analysis of flexible motion in terms of the distance $d_{cc}$ between active sites, Supplemental **Figure S4** plots the evolution of $d_{cc}$ through flexible motion in modes $m_7$-$m_{11}$ for a range of values of the parameter $E_{cut}$. Lower values of $E_{cut}$ correspond to greater limitations on flexible motion and, as expected, they lead to more restricted motion, as measured by this parameter.

The histogram of the MD inter-site distances (**Figure S5**) is derived from the data in Figure 3b and confirms the wide spread of distances with the most common distance being close to 40 Å. Although this inter-domain distance varies widely through the MD simulation, intra-domain distances (exemplified by the distance between the active-site residue Cys61 and a buried residue within the same domain (Cys90)) remain essentially constant (**Figure 3b**) and shows a very tight distribution around 11 Å (**Figure S5**).

**Figure S6** shows that in the majority of modes, the intra-domain RMSD is greatest for the **a'** domain and smallest for **a** and **b** domains. This trend in intra-domain motion is also evident in the analysis of variation of backbone pseudodihedral angles (**Figure 4a**) where both MD simulation and flexibility analysis show greatest variation in the **a'** domain.

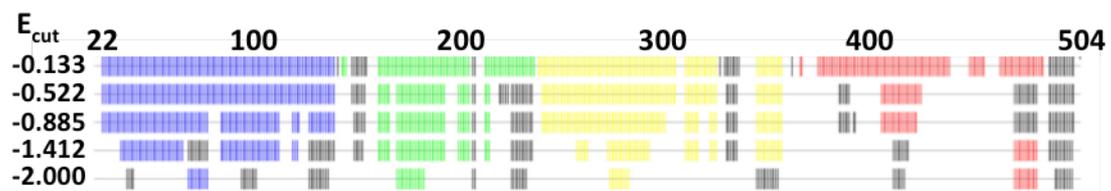

**Figure S1** Rigid cluster decomposition graph. The horizontal axis represents the protein backbone and the vertical axis the energy $E_{cut}$. The residues belonging to rigid clusters are colored as in **Figure 1** whereas the flexible regions are shown as horizontal thin black lines.

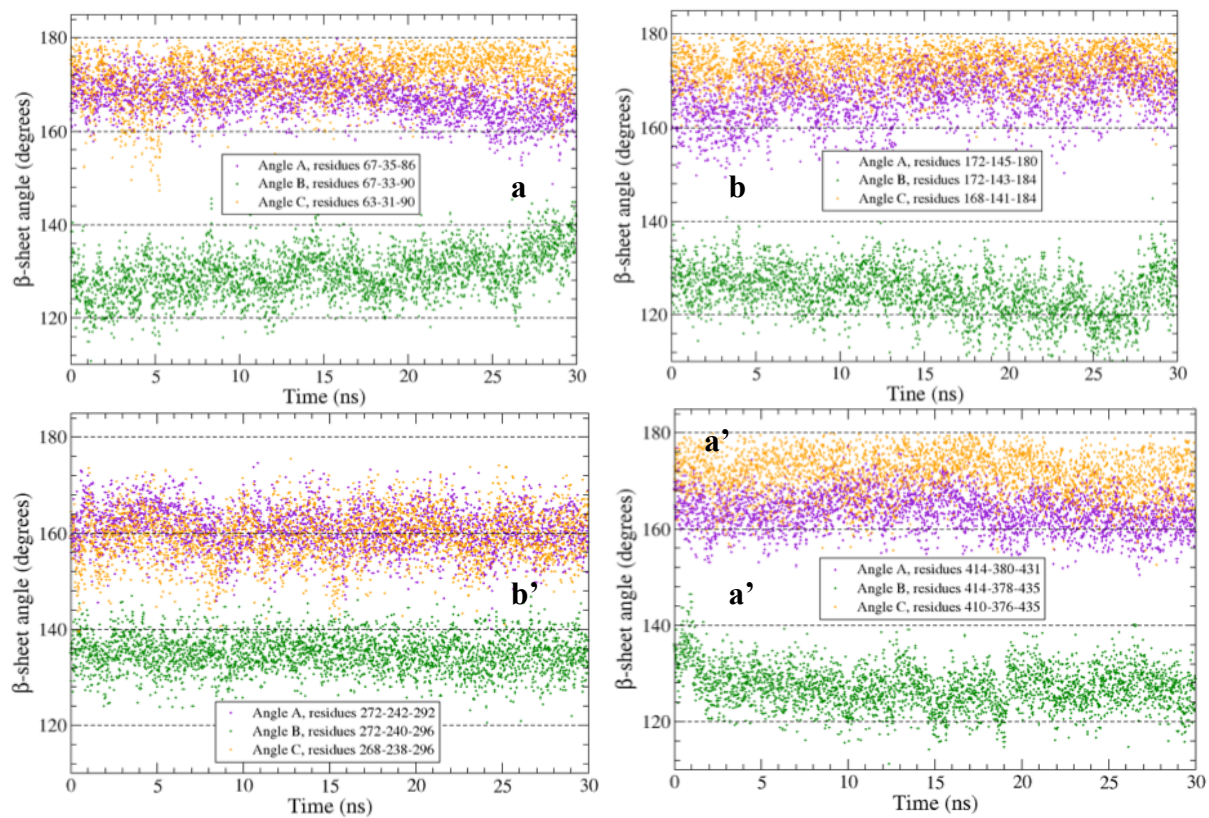

**Figure S2** The $\beta$-sheet angles remain stable along the 30 ns MD trajectory. The three colours correspond to the three angles defined on the indicated residues for each *β*-sheet (cf. **Error! Reference source not found.**c). See also Figure **Figure 1**c.

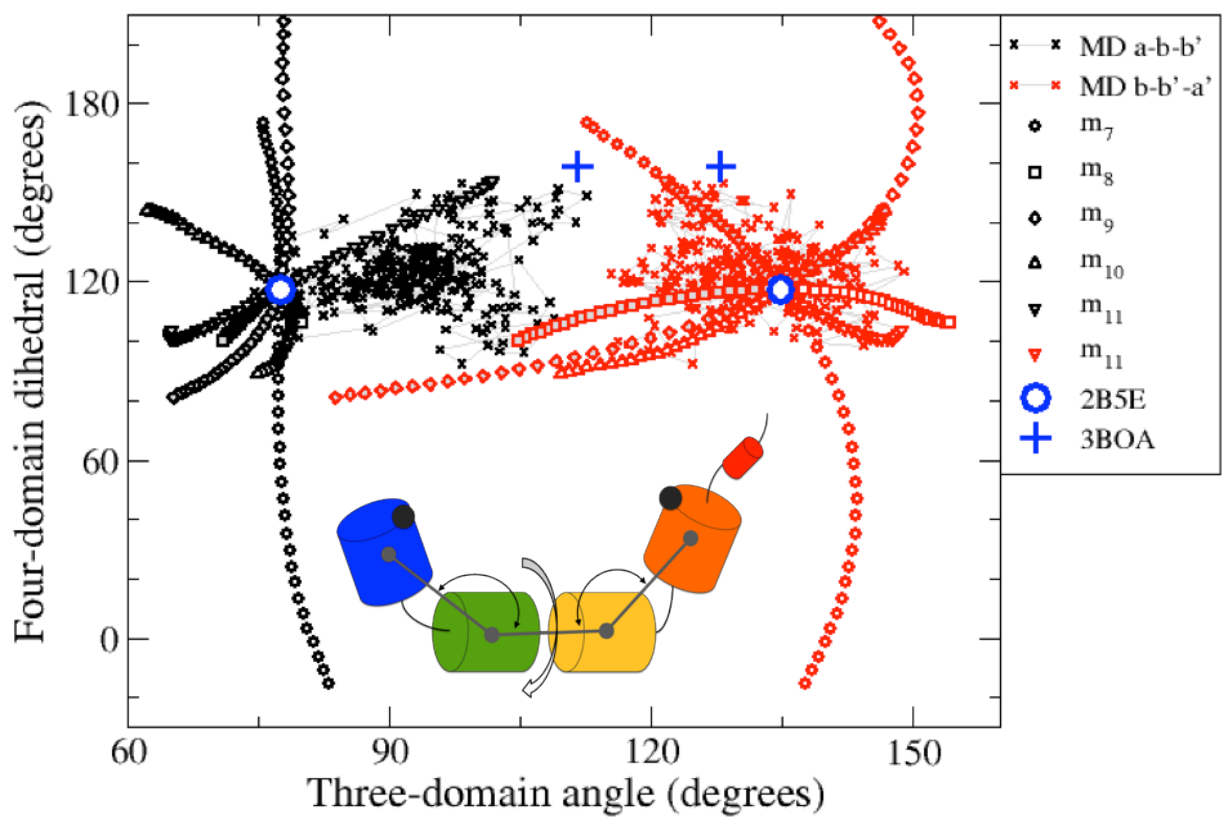

**Figure S3** Four-domain arrangement in PDI. The black (red) data points (× for MD data, others for flexibility modes) show the variation of the four-domain dihedral angle with respect to the **a-b-b'** (**b-b'-a'**) three-domain angle. The open symbols show negative modes $m_{7-}, \ldots, m_{11-}$, the filled symbols are for positive modes $m_{7+}, \ldots, m_{11+}$. The blue circle indicates the starting values for structure 2B5E while the blue cross shows the alternative 3B0A structure. The site chosen to represent each domain is the $C_\alpha$ of a centrally located residue within the β-sheet: the selected residues are ALA 86 (domain a), VAL 191 (domain b), VAL 291 (domain b') and ALA 433 (a'). Inset: Schematic view of the four-domain dihedral arrangement (white arrow) and the 2 three-domain angles (black arrows).

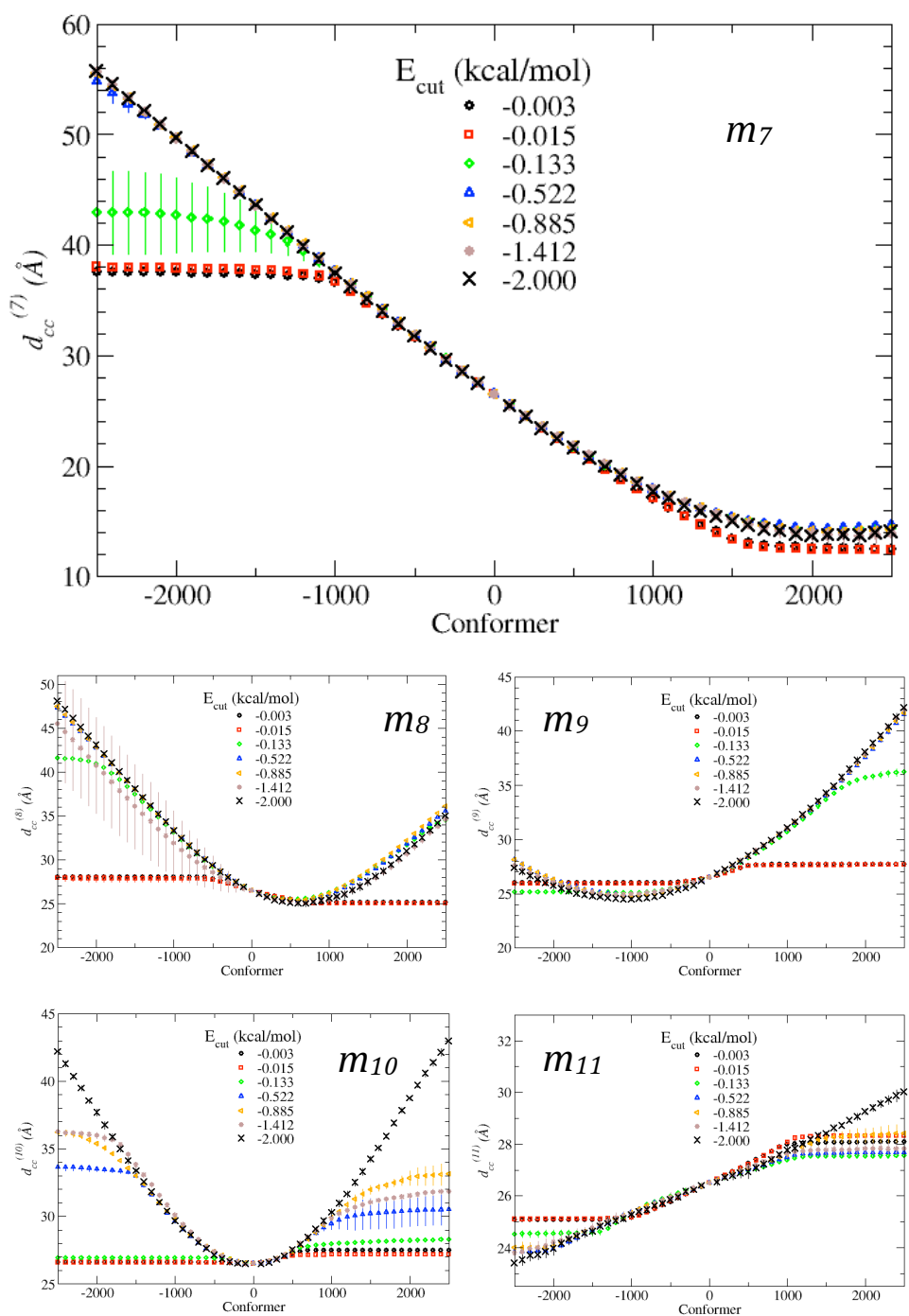

**Figure S4** Distance $d_{cc}$ between the cysteine active sites in the **a** and **a'** domains for various modes ($m_7$ (top), $m_8$ ... $m_{11}$) and $E_{cut}$ values. The $d_{cc}$ distances correspond to the intercysteine distance of the conformers obtained as the protein is projected along (a) mode $m_8$, (b) mode $m_9$, (c) mode $m_{10}$ and (d) mode $m_{11}$ along closing (positive) and opening (negative) directions.

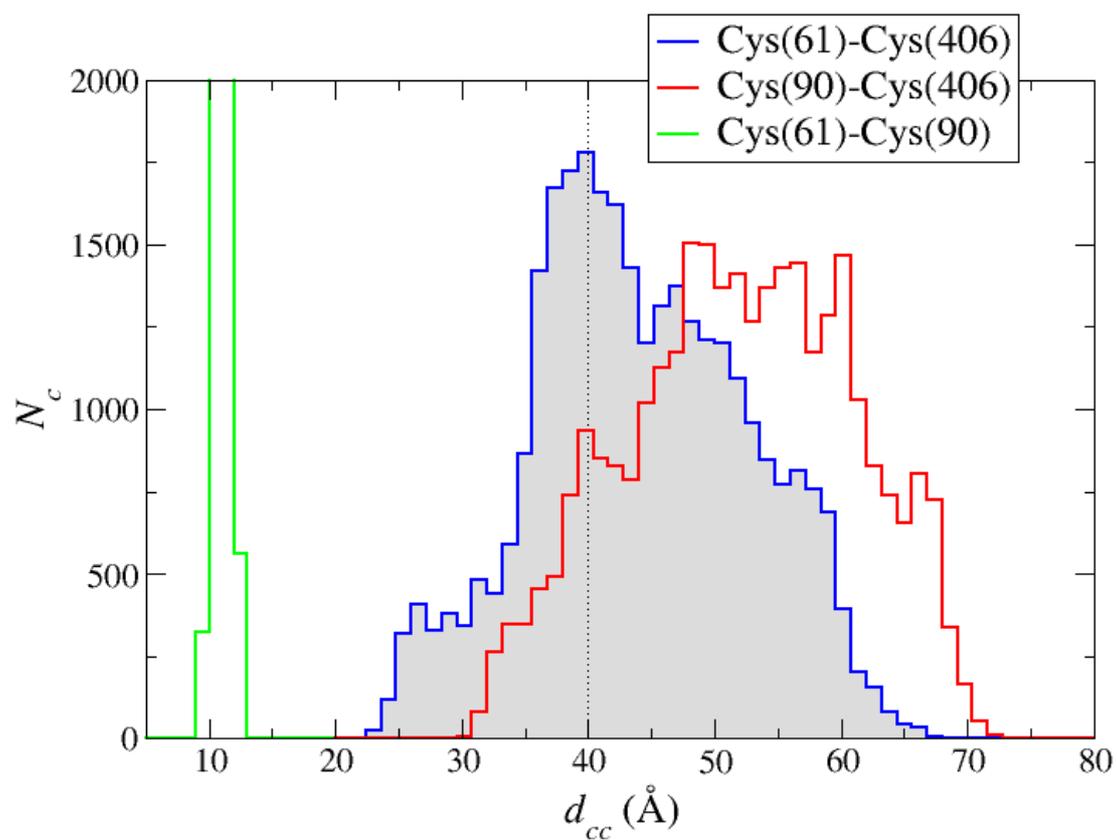

**Figure S5** Histogram of inter-cysteine distances generated through the 30 ns MD simulation. The vertical dotted line indicates the value $d_{cc}$ =40 Å. See also **Figure 3**.

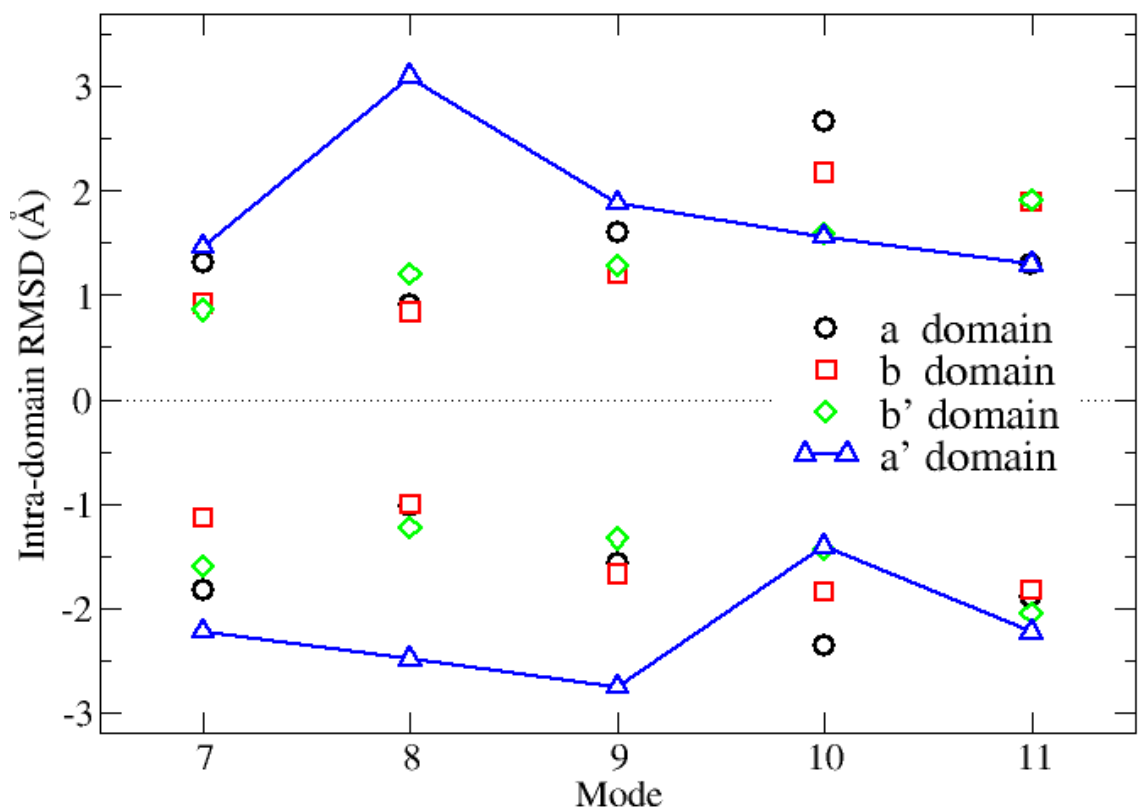

**Figure S6** Change in intra-domain RMSD for modes $m_7, \ldots, m_{11}$. The line connects RMSD values for the **a'** domain, which exhibits the largest intra-domain RMSD for modes 7, 8 and 9. See also **Figure 4**.